\documentclass[aps,preprint,prl,groupedaddress,showpacs]{revtex4-1}
\usepackage[latin9]{inputenc}
\usepackage{textcomp}
\usepackage{amstext}
\usepackage{graphicx}
\usepackage{hyperref}
\usepackage[dvipsnames]{xcolor}
\usepackage{cleveref}
\newcommand\myshade{85}

\colorlet{mylinkcolor}{violet}
\colorlet{mycitecolor}{YellowOrange}
\colorlet{myurlcolor}{Aquamarine}

\hypersetup{
  linkcolor  = mylinkcolor!\myshade!black,
  citecolor  = mycitecolor!\myshade!black,
  urlcolor   = myurlcolor!\myshade!black,
  colorlinks = true,
}
\makeatletter

\newcommand{\lyxmathsym}[1]{\ifmmode\begingroup\def\b@ld{bold}
  \text{\ifx\math@version\b@ld\bfseries\fi#1}\endgroup\else#1\fi}

\usepackage{float}
\usepackage[normalem]{ulem}
\usepackage{color}

\newcommand{\ket}[1]{|#1\rangle}

\makeatother

\begin{document}
\begin{abstract}
    The Quantum Fourier Transformation ($QFT$) is a key building block for a whole wealth of quantum algorithms.
    Despite its proven efficiency, only a few proof-of-principle demonstrations have been reported. 
    Here we utilize $QFT$ to enhance the performance of a quantum sensor. 
    We implement the $QFT$ algorithm in a hybrid quantum register consisting of a nitrogen-vacancy (NV) center electron spin and three nuclear spins.
    The $QFT$ runs on the nuclear spins and serves to process the sensor - NV electron spin signal. 
    We demonstrate $QFT$ for quantum (spins) and classical signals (radio frequency (RF) ) with near Heisenberg limited precision scaling. 
    We further show the application of $QFT$ for demultiplexing the nuclear magnetic resonance (NMR) signal of two distinct target nuclear spins. 
    Our results mark the application of a complex quantum algorithm in sensing which is of particular interest for high dynamic range quantum sensing and nanoscale NMR spectroscopy experiments. 
\end{abstract}

\title{Quantum Fourier transform for quantum sensing}

\author{Vadim Vorobyov$^{1,*}$, Sebastian Zaiser$^{1}$, Nikolas Abt$^{1}$, Jonas Meinel$^{1}$, Durga Dasari$^{1}$, Philipp Neumann$^{1}$ and J{\"o}rg Wrachtrup$^{1}$}
\email{v.vorobyov@pi3.uni-stuttgart.de, j.wrachtrup@pi3.uni-stuttgart.de}
\affiliation{$^{1}$ 3. Physikalisches Institut, IQST and Centre for Applied Quantum Technologies, and MPI
for Solid State Research, University of Stuttgart, Pfaffenwaldring
57, 70569 Stuttgart, Germany}
\maketitle

\section*{Introduction}
The Quantum Fourier transform ($QFT$) is a key element in a variety of  quantum algorithms, such as conventional phase estimation, Shor's prime
factorization protocol \cite{Shor.1994}, period and order finding \cite{Nielsen.2010}, or in the modern applications of quantum machine learning \cite{MLreview2017}. 
Along with the large theoretical research triggered in the fields of physics, chemistry and computer sciences over the past two decades, there have been experimental demonstrations of the algorithm using various physical systems involving superconducting qubits \cite{GarciaMartin.2018},
trapped ions \cite{IONS_QFT_2016}, nuclear magnetic resonance (NMR) \cite{Dory_NMR_QFT_2001}, and integrated optics \cite{Optical_Circuit_review_2019}. 
Yet, besides these proof-of-principle demonstrations no practical use of the $QFT$ in e.g. factorization of large numbers has been made, because it would require a so far not available number of qubits.

 On the other hand, classical Fourier transformation is used in a much larger variety of applications than the ones highlighted above for QFT. Excellent examples are found in signal analysis, specifically in spectroscopy, where in e.g. infrared and particularly in nuclear magnetic resonance, fast Fourier transformation (FFT) is the hallmark of modern biomolecular nuclear magnetic resonance structure analysis. In this technique signal post processing is used to demultiplex a complex multifrequency signal essentially yielding a massive reduction in signal acquisition time over measuring each frequency separately. Here we show, that using a multiqubit sensor and QFT simultaneously can demultiplex a multifrequency signal in situ and yield superior sensor performance.     
 
In a typical quantum sensing scenario a physical quantity $\alpha$ which gives rise to the energy shift $\delta E(\alpha)$ and leads to the phase accumulation $\phi = \delta E(\alpha)\cdot \tau $ of the quantum state used for the measurement. This phase shift is measured by e.g. Ramsey interferometry \cite{Ramsey1950}. 
The measurement sensitivity scales with the phase acquisition time $\tau$ as $1/\sqrt{\tau}$ in the standard quantum limit. 
In case multiple $\delta E (\alpha_i)$ are detected, for example caused by a complex nuclear magnetic resonance spectrum, the acquired phase would 
comprise multiple phases $\phi_i$. To yield an unambiguous phase readout,  the phase evolution is restricted to the interval $[ 0, \pi ]$, when using a single sensor spin for phase measurements. 
Since the acquired phase $\phi = \sum_i \phi_i = \tau \sum \delta E(\alpha_i)$, is given by the evolution time $\tau$ multiplied with the sum of all frequency components, the maximum acquisition time is thus determined by the largest spectral component, significantly reducing the overall sensitivity. 
The addition of multiple memory and register qubits and use of $QFT$ allows to extend the dynamic range and hence sensitivity in such measurement and to demultiplex the individual harmonics from individual sources (spins) onto various memories outputs as depicted on Fig. \ref{fig:Fig1}.

We implement the $QFT$ algorithm on a three nuclear ancillary spin register coupled to nitrogen vacancy (NV) center in diamond and demonstrate phase digitization. Further we use this to perform high dynamic range sensing of a RF signal and parallel correlation spectroscopy of two distinct weakly coupled target spins.
We show here, that extending the dynamic range of the sensor by encoding the acquired phase $\phi$ into memory states $|m_1m_2..\rangle $, and using the $QFT$ algorithm to demultiplex the signal harmonics also enhances the overall sensitivity of the sensor.

\section*{Results}

\subsection*{Implementation of QFT algorithm}
\begin{figure}
\includegraphics[width=1.0\textwidth]{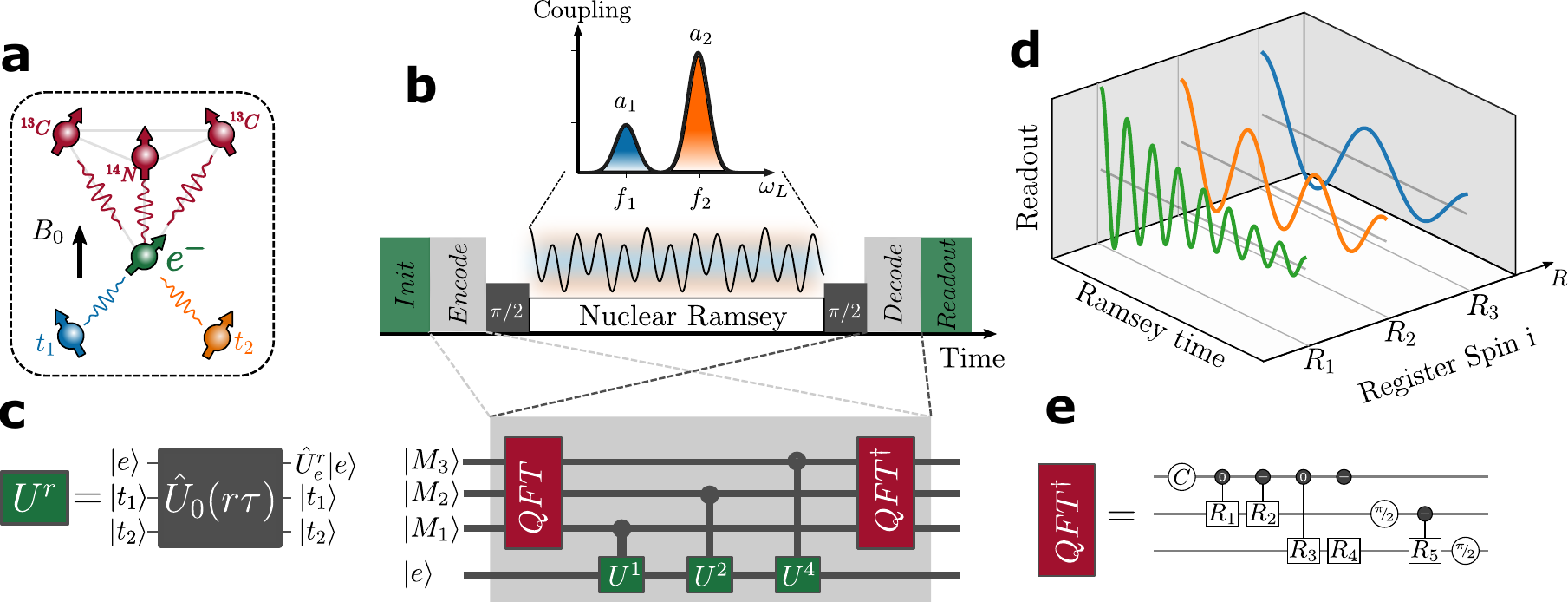}
\caption{(a) Schematic representation of the sensor consisting of a single sensor spin (green)
which collects phase information of a distant target spin (blue and orange) and distributes it onto a local qubit 
register (red) made of one qutrit ($^{14}N$ nuclear spin) and two qubits ($^{13}C$ nuclear spins).
(b) The target system spectrum, consisting of multiple frequencies $f_1$, $f_2$,... with corresponding amplitudes $a_1$, $a_2$,...
with a schematic circuit for sensing of multiple target spins with implementation of $QFT$ and $QFT^{\dag}$. (c) The phase acquisition unitary gate for sensing of target nuclear spins $t_1$ and $t_2$. 
(d) The readout result of the register qubits after performing measurements with the $QFT$ algorithm. Each memory stores Ramsey oscillation of individual target nuclear spins. (e) $QFT^{\dag}$ quantum circuit for our hybrid quantum register}
\label{fig:Fig1}
\end{figure}

Our $QFT$ algorithm utilizes an electron spin as sensor to acquire a phase $\phi$ which we subsequently transfer onto memory and processing qubits 
to perform a $QFT$, the result of which is read out after the phase acquisition is accomplished \cite{Nielsen.2010}. 
To implement our experiment we use a single electron spin of an NV center in diamond with long coherence time at room temperature as a probe qubit capable of sensing various external quantities 
\cite{Dolde.2011,Acosta.2010b,taylor_high_sensitive}. 
As memory register we use three well isolated and thus long-lived individual nearby nuclear spins ($^{14}N$, $^{13}C_1$,$^{13}C_2$) which form a twelve 
level quantum system for storage and processing of the sensed information as schematically depicted in Fig. \ref{fig:Fig1}a. 
We note that we can perform single-shot readout on all three nuclear spins. 
The electron spin sensor measures small magnetic fields and distributes the phase acquired during the sensing step $U^{2^i}$ to the $i^{th}$ memory qubit (see Fig. \ref{fig:Fig1}. (b), (c)). 
The long lifetime of the memories allows for long phase storage and consequently high spectral resolution due to large correlation times. 
In our setting,  $\alpha$ is a magnetic field created by proximal target spins or a classical signal, or for example nuclear spins of an unknown complex molecule or spin cluster. 
In the case of a nuclear spin-bath, the multi-frequency signal from the sample with peak frequencies $f_1$ and $f_2$ and amplitudes $a_1$ and $a_2$ results in a
beating in the free precession signal of the nuclear spins. The task at hand is to demultiplex this by applying Quantum Fourier transformation. 

In case of our hybrid qubit-qutrit register the $QFT$ is similar to a standard register composed of qubits i.e. realized with use of the Hadamard and control rotation gates. 
Fig. \ref{fig:Fig1}e shows a circuit representation of a $QFT$ and $QFT^{\dag}$  algorithm for an effective twelve level system, consisting of one qutrit and two qubits (see Supplementary).
In general, the $QFT$ involves local Hadamard (Chrestenson \cite{AlRabadi.2002}) gates for qubits (qutrit) and a large 
number $(O(n^{2}))$ of conditional non-local rotational gates, and are implemented using optimal control (see Supplementary) to enhance fidelity. 
 
In the version of the algorithm adapted to our hybrid qubit-qutrit quantum register (see Fig. \ref{fig:Fig1}c) the register spins are initialized in a superposition initial state $\ket{+}$ with zero phase using local Hadamard and Chrestenson gates. After that, the sensor interacts with the target system, attains phase information and stores it onto the register spins using controlled $U^i$ gates \cite{wald}. The phase acquisition gate $\hat{U}$ in Fig. \ref{fig:Fig1}b,c presents a unitary operator  acting on the electron spin for example by $\hat{U}_e=exp\{i\sum_{t_i} A_{zz}^i S_z I_z^i \tau\}$ or $\hat{U}_e=exp\{i\sum_{t_i} A_{zx}^i S_z I_x^i \tau\}$, depending on the control sequence and on the state of the target spin system. 
Finally, $QFT^{\dag}$ transforms the acquired phases to a bit representation (digitization) by mapping it onto populations of the nuclear spin register which are then either used as a classical memory during the correlation time or read out 
through a single-shot measurement \cite{Neumann.2010b}. 

In a classical Fourier transformation of e.g. an NMR signal, the data is acquired first and then the Fourier transformation is applied subsequently. Applying the QFT however, requires a simultaneous recording of data and performing $QFT^{\dag}$. 
We start by first applying the $QFT$ on the initialized nuclear quantum register and then transfer the measured phases from the electron spin to the nuclear quantum bits.
The role of $QFT^{\dag}$ on the nuclear quantum register is to make the mapping of acquired phases during the phase acquisition more efficient by realizing an unambiguous digitization of the phase. To demonstrate the efficiency of this step and compare $QFT$ with other phase conversion methods, we first prepare the initial state of the register with a certain phase $\phi$ mimicking a phase acquired by the sensor spin (see Fig. \ref{fig:Fig2})(see Supplementary). The figure shows the result of a readout of the nuclear register (i.e. its bit values) for different input phases.  
The simplest way to convert this phase into a detectable $I_z$ magnetization are Hadamard gates on each nuclear spin which rotate the spin from the phase plane into the $z$ direction. For comparison we show the result of the same initial state of the nuclear spins with the $QFT^{\dag}$  protocol. As apparent from the upper row of Fig. \ref{fig:Fig2}. b,c  the application of local gates results in quantum register readouts scattered throughout the whole logical space. Most importantly, most phases (except $\phi=\pi$) result in multiple bit values, i.e. the register readout does not result in an unambiguous phase. On top, the readout contrast is reduced by $1/a$, when $a$ is the number of bit value results per given phase. 
However, the contrast is maintained upon application of the $QFT^\dag{}$, and the measurement output corresponds to a well defined phase. In Fig. \ref{fig:Fig2}b,c  the analytical and experimental result for the application of $H^{\otimes3}$ and $QFT^{\dag}$ on the same initial phase encoded state is shown respectively. Besides the excellent agreement between experiment and theory the results show excellent fidelity of our $QFT$ based protocol for sensing with hybrid quantum devices containing multi-qubit registers (see Supplementary).  For the $QFT$ based protocol, the fidelity stays constant over the whole range of $\phi$, and the individual qubits are projected onto their eigen basis, minimizing the spin projection noise during the projective readout of the register state. 
A further advantage of the $QFT$ is that in previous application of in-situ correlation 
 
In classical Fourier transform NMR, a multifrequency signal is recorded in a Ramsey-type experiment and its frequency components are subsequently determined by a Fourier transformation. 
In our implementation we measure an oscillating magnetic field and process the $QFT$ in-situ (see Fig. \ref{fig:Fig3}a). 
A notable difference between the classical approach and our quantum sensor is that we first need to convert the frequency of the oscillating field into a phase which can be transferred to the nuclear quantum register. 
To accomplish this, 
we design and implement phase gates $U^i$, converting the AC magnetic field into a phase acquired by the electron spin of NV center and imprint it onto various nuclear register qubits \cite{Zaiser.2016}.
Essentially, $U^i$ comprises a train of phase inversion pulses which are commensurate with the oscillation of the AC magnetic field sandwiched between two CNOT gates which write the phase on the nuclear register (see Fig. \ref{fig:Fig3}b). 
Our measurement starts with memory initialization, followed by a sensing stage. In the sensing stage, the sensor qubit attains phase information related to the incident magnetic field and swaps it onto the memory qubits where it is stored in form of a relative phase and is transformed through the $QFT^{\dag}$ into a binary representation, which is then finally read out.
Fig. \ref{fig:Fig3}c shows the output of the $QFT^{\dag}$ in the register states basis representation $|m_1m_2m_3\rangle$.
Analogous to Fig. \ref{fig:Fig2}c, the image shows the unambiguous representation of the field amplitude with respect of the measurement result output, which corresponds to full $2\pi$ phase estimation range on the twelve-level quantum register. 
We would like to highlight here the enhanced dynamic range of the QFT alongside the precision of the sensing protocol. As the DR scales with the dimension of the Hilbert space (see Supp. for details) we experimentally find a $\sim 12 x$ improvement when compared to the single qubit case.
To see the benefit of using a $QFT$ based phase estimation we analyse the Fisher information in 
Fig. \ref{fig:Fig3}.d and use it to visualize the scaling behavior of our phase estimation 
scheme with respect to the number of qubit resources used.
To do this we analyze the output state in the sensing protocol as a function of the acquired phase $\phi$, using the expression for the Fisher information for the pure states, viz., 
$F_q = 4(\langle \partial_a \psi | \partial_a \psi \rangle - |\langle \partial_a \psi | \psi \rangle |^2)$. 
We estimate the theoretical Fisher information of the final state of the sensing protocol with respect to parameter $\phi$, and compare it to the expressions obtained for the case of $n$ non-entangled qubits, and $n$ fully entangled qubits (NOON states), which represents the Standard Quantum limit and Heisenberg scaling limit in number of used qubit resources (see Supplementary). 
In the same plot we show the experimentally achieved values of the Fisher information for the case of a twelve-level register. These values were estimated for the experiment shown in Fig. \ref{fig:Fig2} and Fig. \ref{fig:Fig3}  for the bare $QFT^{\dag}$ and $QFT^{\dag}$ with sensing. While the experimental values show a slight deviation in the achievable information compared to the theoretically calculated case, they outperform the Standard Quantum Limit and approach  Heisenberg scaling (see Methods).

Finally we turn to measuring signals of multiple, non-identical nuclear spins.
State-of-the-art quantum sensing protocols with NV centers in diamond enable high resolution nuclear magnetic resonance experiments reaching sub Hz resolution.  
This is either accomplished by measuring the signal in subsequent measurements and correlating them afterwards \cite{Schmitt.2017,Glenn.2018}, or by in-situ correlation spectroscopy \cite{Laraoui.2013, Zaiser.2016}. 
In the latter first the oscillating magnetic field caused by the Larmor precession of the target nuclei is converted into a phase, similar to the above detection of the test field. However, the phase detection sequence is repeated after a certain (correlation) time $T_c$ to achieve a higher frequency resolution \cite{Aslam.2018}. 


Here we extend this method by a multi-qubit memory and apply $QFT$ on these qubits. 
The general working principle of this method is depicted schematically in Fig. \ref{fig:Fig4}a. 
It is using multiple nuclear spins to store the phase of the electron spin and in addition is applying $QFT$ to the attained data.
The protocol starts with applying a $QFT$ to the initialised nuclear spin register. Subsequently the NMR signal of the target spins is measured by the electron spin and its phase information is encoded in the nuclear spin register. 
A $QFT$ algorithm is then applied to these nuclear spin quantum states, essentially encoding the phase information into nuclear spin eigenstates for long time storage. 
Arbitrary operations could be done on the target spins during that period, for example a Ramsey sequence or a single $\pi$ pulse, before the memory is mapped back for the second correlation step. 
The operations on the target nuclear spins change the local magnetic field of the sensor electron spin. In the final decoding stage of the algorithm, this new magnetic field is compared (correlated) with the one measured during the encoding state.  
The addition of multiple register qubits is a natural choice for expanding the memory capacity and hence improve the function of the method. However, a proper way of processing the information is required  and this is achieved by the $QFT$. 

During the protocol, the $QFT$ is used to demultiplex the signals originating from the mixture of target spins onto separate memories outputs. 
To demonstrate this we choose weakly coupled nuclear spins of $^{13}C$ with $A_{zz}$ coupling of 6 kHz ($t_1$) and 12.4 kHz ($t_2$) as two target spins.

The $QFT$ and $QFT^{\dag}$ in between the two sensing steps and the duration of interrogation time $\tau$ during phase accumulation was chosen such that the signal originating from the 12 kHz coupled target spin is reflected by the result measured on $^{13}C_{414}$ nuclear spin memory, whereas the signal originated by 6 kHz target spin is measured by the output of the another register qubit  ($^{13}C_{90}$)(see Supplementary). 

We perform correlation spectroscopy of these target spins with a Ramsey type 
measurement in between the sensing steps. 
For the nuclear spin Ramsey $\pi/2$ pulses, a frequency detuning of 2.5 kHz to the weaker coupled target spin was chosen, such that we should measure an oscillation with 2.5 kHz on $^{13}C_{90}$ and an oscillation of 3.8 kHz on the $^{13}C_{414}$ memory output, respectively. This is observed in the measurement and depicted in Fig. \ref{fig:Fig4}e-f achieving an NMR linewidth of 76 Hz and 68 Hz. 

In our measurement protocol the phase acquisition is adjusted such, that the least significant qubit (LSQ) acquires a total phase $\phi_{LSQ}=4\cdot 2 \tau A_{t_2}=2\pi$ and the most significant qubit (MSQ) a phase of $\phi_{MSQ}=4\cdot \tau A_{t_2}=\pi$  as the stronger coupled target $t_2$ spin flips during the correlation time. The register acquires a phase $\phi_{LSQ}=4\cdot 2 \tau A_{t_1}=\pi$ on the LSQ and $\phi_{MSQ}=4\cdot \tau A_{t_1}= 0  (mod 2\pi)$ on the MSQ when the weaker coupled target spin $t_1$ flips. Hence, we pick a sensing time $\tau$, such that it fits the energy difference of the combined target spins. In our case, the sensing time $\tau$ was chosen to be $\tau = 3\pi/8\cdot(A_{t_1}+A_{t_2})\approx (4\cdot12 kHz)^{-1}$ as depicted in Fig. \ref{fig:Fig4}d.

This setting results in the mapping of phases acquired due to the signal of  the individual target spins as shown in Fig. \ref{fig:Fig4}d onto the register states.
As a result when performing Ramsey type correlation measurement on two target spins their oscillation are directly mapped onto the output populations of memory spins which are single-shot readout in Fig. \ref{fig:Fig4}e. By performing an Fast-Fourier Transform (FFT) on the results of individual memory outputs we show two resonances corresponding to 12 and 6 kHz $^{13}C$ nuclear spins with the resolution of $\approx 70 Hz$  depicted in Fig. \ref{fig:Fig4}f.

In conclusion we demonstrated the first implementation of $QFT$ using individual solid-state nuclear spins in diamond. Combining $QFT$ with sensing we extended the capability of diamond based quantum sensors and realized a multiqubit phase estimation circuit in a correlation spectroscopy measurements, and monitored the dynamics of a two target spin with high precision. 
Their NMR signals were demultiplexed and read out as a separate quantum register outputs.
Our results show, that multiple qubit algorithms can be  beneficial for quantum sensing, even in terms of sensitivity scaling.  

It has been shown, that a NV register can comprise up to seven nuclear qubits, extending the number of frequencies simultaneously detected to a value which is of interest to be applied in multi species nanoscale NMR.


\section*{Methods}

\subsection*{NV center - nuclear spin system}
In our experiments we use a single electron
spin of an individual NV center as a sensor
The electron spin has a long $T_2$ coherence time
\cite{Zaiser.2016} at ambient conditions despite its large
coupling to external fields and consequently a high intrinsic sensitivity. 
To perform the $QFT$, one needs highly controllable nuclear spins that constitute the memory register. 
The NV center has several hyperfine coupled nuclear spins in the environment having much longer coherence times as compared to the electron spin but having a low susceptibility to the environment. 
The NV center has several hyperfine coupled nuclear spins in the environment having much longer coherence times as compared to the electron spin but having a low susceptibility to the environment. 
The co-processor is formed using the strongly coupled
(2.16 MHz) $^{14}N$ nuclear spin of the NV center, and two $^{13}C$ nuclear
spins (labelled as $^{13}C_{414}$ and $^{13}C_{90}$) with hyperfine
coupling along the quantization axis of the NV center $A_{zz}\approx414 kHz$
and $A_{zz}\approx90kHz$ respectively \cite{Waldherr.2014}.
It presents an effective 12-level system and allows to run a general set of quantum operations including the
QFT algorithm. 
Additionally, our register allows for single-shot readout of nuclear spins and thus efficient extraction of sensing
information \cite{Neumann.2010b, Waldherr.2014}. 
Due to the relatively strong hyperfine coupling we can resolve all the twelve lines in the ODMR spectrum of the NV center which has an $T_2^*$ time about 20 $\mu s$.
This marks the key step of our protocol, as the digitized phase is encoded
as populations among the 12-levels with a resolution of $\phi/12$. 
The individual addressing of these levels set the limit on our precision. 
Using strongly coupled nuclear spins as memory qubits allows them to be individually addressed and readout.
(see Supplementary Information).

\subsection*{Precision and dynamic range scaling}
\subsubsection*{}
For the multiple independent qubits the Quantum Cramer-Rao Bound (QCRB) gives the ultimate bound for precision of a parameter estimation.
The formula for the QCRB reads $\Delta \phi_{SQL} = \frac{1}{\sqrt{N_m}\sqrt{F_q}} =\frac{1}{\sqrt{N_m}\sqrt{n}}$, where $N_m$ is the number of measurements and $n$ is the number of independent qubits. 
It is also worth noting that in a sensing scenario, when an unknown parameter $\alpha$ has to be determined, one can write that $\phi = \delta E(\alpha) \tau / h$. 
The corresponding final state reads as $|\psi\rangle = \frac{1}{\sqrt{2}} (|0\rangle + e^{i \frac{\partial E}{\partial \alpha} \delta \alpha \tau /h} |1\rangle)$, and the Fisher Information with respect to the parameter $\alpha$ reads $F_{\alpha}^q = n \partial E_{\alpha}^2 \tau^2$. 
$\Delta \alpha = h \frac{1}{\sqrt{N_m}}\frac{1}{\sqrt{n}\partial_\alpha E \tau }$ comprises the Standard Quantum limit in number of measurements and Heisenberg limit scaling in measurement time. 
It shows SQL limit scaling of precision in the number of independent qubits. 
When multiple independent qubits are subject to a phase estimation algorithm, the Fisher information in phase estimation reads $F_q = \sum_{j=0}^{n-1} 2^{2j}=1/3 (4^n-1)$, the QCRB for phase $\Delta \phi_{PEA} = \frac{1}{\sqrt{N_m}}\frac{1}{\sqrt{(4^n-1)/3}} \approx \sqrt{\frac{3}{N_m}}\frac{1}{2^n}$ at large n and it reduces exponentially with number of used qubits. 
On the other side, when a phase is acquired due to the sensing, then $\phi = \tau \partial_\alpha E \delta \alpha /h$. In that case $F_\alpha^q = \sum_{j=0}^{n-1}(2^{j}\tau \partial_\alpha E/h)^2$, and $\Delta \alpha = h\sqrt{\frac{3}{N_m}}\frac{1}{2^n}\frac{1}{ \tau \partial_\alpha E}$.
Since the total time of the measurement is approximately $T\approx 2^n \tau$ then $\Delta \alpha = h\sqrt{\frac{3}{N_m}}\frac{1}{T \partial_\alpha E}$.
In comparison, for the case of n entangled qubits as an initial state, when a similar phase estimation operation is applied, the Fisher information reads $F_q = (\sum_{j=0}^{n-1}2^i)^2 = (2^n-1)^2$. 
In the case of sensing a parameter $\alpha$ as in the previous case,  $F_{\alpha}^q = (\sum_{j=0}^{n-1}2^i \tau \partial_\alpha E)^2$, and $\Delta \phi_{HL} = \frac{1}{\sqrt{N_m}}\frac{1}{(2^n-1)}$, which has an exponential scaling in number of qubits and for high n scales as for non-entangled qubits.
The precision of sensing of a parameter is equal to $\Delta \alpha = h \frac{1}{\sqrt{N_m}}\frac{1}{(2^n-1)}\frac{1}{\tau \partial_\alpha E}$. 
To summarize the QCRB gives the following scaling for the precision of conventional sensing using n non-interacting qubits, QPEA, and QPEA+Entangled state: 
\begin{equation}
   \Delta \alpha_{SQL} = h\frac{1}{\sqrt{N_m}}\frac{1}{\tau \partial_\alpha E}
\end{equation}
\begin{equation}
    \Delta \alpha_{QPEA} = h\sqrt{\frac{3}{N_m (4^n-1)}}\frac{1}{\tau \partial_\alpha E}
\end{equation}
\begin{equation}
    \Delta \alpha = h \frac{1}{\sqrt{N_m}}\frac{1}{(2^n-1)}\frac{1}{\tau \partial_\alpha E}
\end{equation}

\subsubsection*{Dynamic range}
In a quantum measurement the $2\pi$ periodicity of the acquired phase limits the measurement range $R_\alpha  =\pi/(\partial_\alpha E \tau)$ over which the parameter $\alpha$ could be unambiguously determined. Thus there is a trade-off between the precision ($\Delta$) and maximum range of the measurement ($R$), efficiently represented via a dynamic range ($DR=R/\Delta$). In a standard case of n qubit sensing using Ramsey type sequence $DR_{SQL} = \frac{\pi}{\tau\partial_\alpha E} / h\frac{1}{\sqrt{N_m}}\frac{1}{\tau \partial_\alpha E} = \frac{\pi }{h}\sqrt{N_m}$. 

In the case of phase estimation algorithm with a $QFT$ the maximum range is given by the smallest phase acquisition time $\tau$ and phase could be resolved in whole $0-2\pi$ range, while the precision is given by the longest phase acquisition time $\tau\cdot 2^n$. In that case $DR_{QPEA} = \frac{2\pi}{\tau \partial_\alpha E} / h\sqrt{\frac{3}{N_m (4^n-1)}}\frac{1}{\tau \partial_\alpha E}= \frac{2\pi}{\sqrt{3}h} \sqrt{N_m} \sqrt{(4^n-1)} \approx 2/\sqrt{3} DR_{SQL} \times 2^n$, where n is the number of qubits. In summary the phase estimation algorithm and $QFT$ exponentially increase the dynamic range with the number of qubits in use.

\bibliography{references} 
\bibliographystyle{ieeetr}

\textbf{Acknowledgments}
We acknowledge financial support by the German Science Foundation (the DFG) via SPP1601, FOR2724, the European Research Council (ASTERIQS, SMel), the Max Planck Society, the Volkswagen Stiftung

 \textbf{Author Contributions} \\
 PN and JW developed the initial idea of the experiment. 
 NA, SZ, VV, JM, PN and JW conducted the experiment. 
 VV, DD, PN and JW performed the theoretical analysis. 
 All authors contributed to the writing of the manuscripts.
\\
 \textbf{Author Information}

\newpage{}

\newpage{}

\begin{figure}
\includegraphics[width=1.0\textwidth]{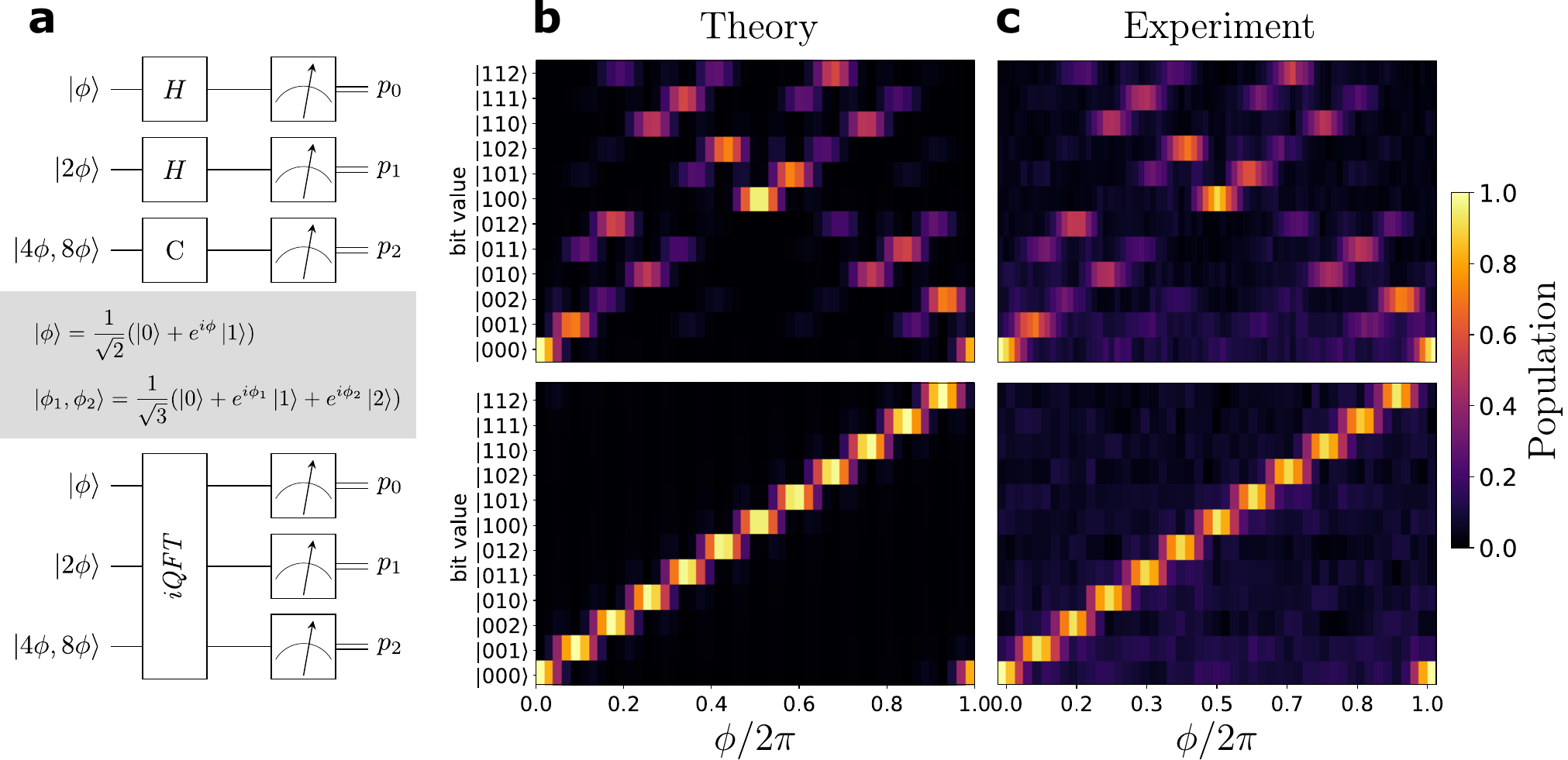} 
\caption{{Measurement outcome of the register after $QFT$.  (a) Quantum circuits of two readout methods of register prepared in arbitrary phase state $|\Psi\rangle=|\phi\rangle \otimes |2\phi\rangle \otimes |4\phi, 8\phi\rangle$. The upper circuit maps of the phase basis onto the population basis using local single qubit (qutrit) Hadamard gates, whereas the lower one uses $QFT^{\dag}$. (b) and (c) are results of theoretical calculations and experimental realization of the circuits on our system. Results of the experiment are in excellent agreement with calculations of the circuit output. Notably in the $QFT^{\dag}$ case there is no ambiguity between regions of phase $0 - \pi$ and $\pi - 2\pi$. Additionally the $QFT^{\dag}$ prepares final states which are close to the eigen states of the register, minimizing the loss of purity due to dephasing. 
}}
\label{fig:Fig2}
\end{figure}


\newpage{}
\begin{figure}
\includegraphics[width=1.0\textwidth]{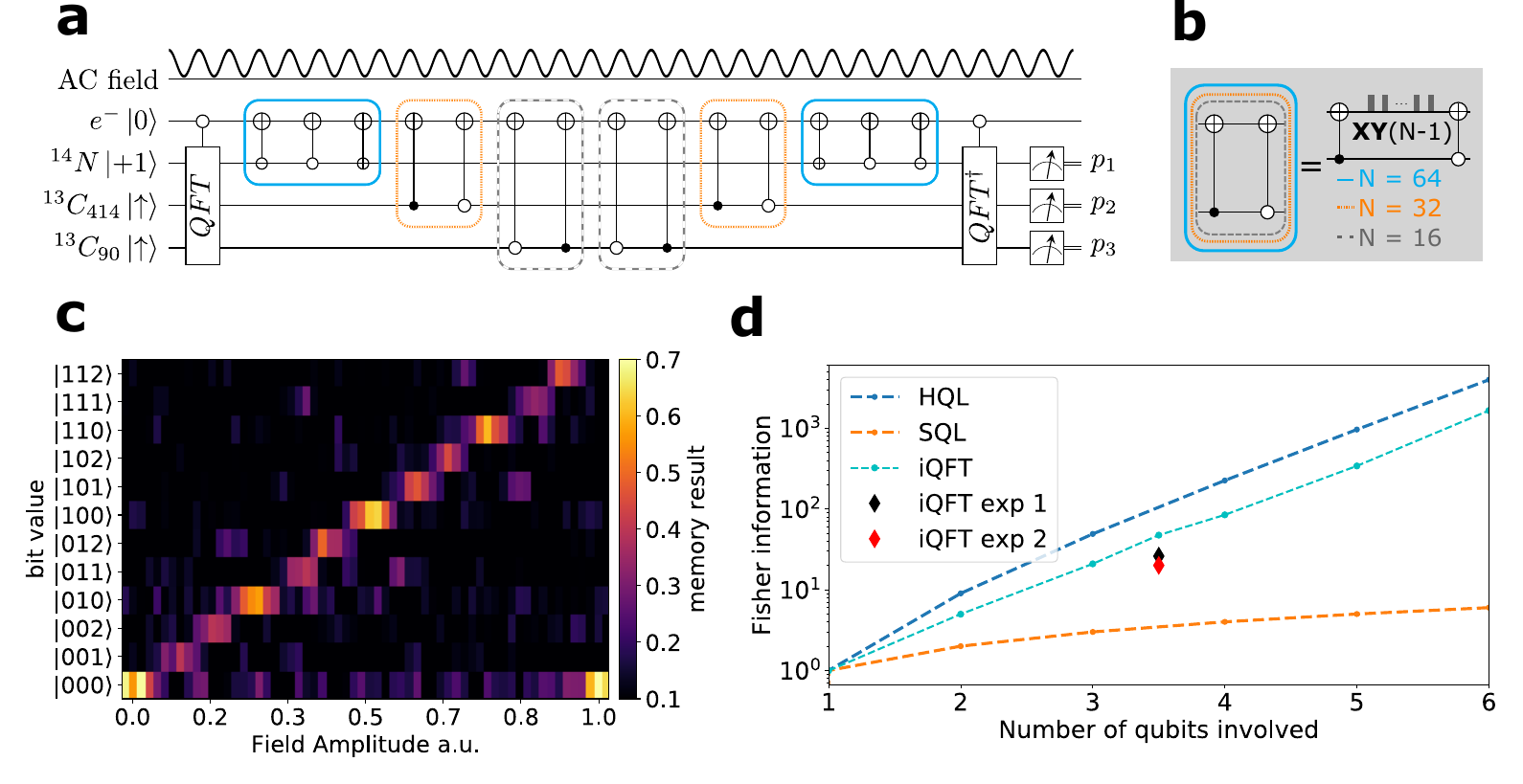} \caption{{
(a) Sensing of an external RF field with the $QFT$. Usage of multiple register qubits allows to enhance the dynamic range of the sensor using multiple interrogation times, 
(b) Schematic representation of the quantum circuit which serves for detection of the artificial field
(c) Measurement outcome of the memory registers 
(d) Fisher information estimated for the case of the inverse Quantum Fourier Transform ($QFT^{\dag}$) sensing protocol, the Heisenberg scaling limit and the Standard Quantum Limit (SQL). Cyan points connected by the thin dashed line marks a simulation of the $QFT^{\dag}$ on the n-qubit circuit. Black and Red diamond are experimental results of the bare $QFT^{\dag}$ algorithm and the $QFT^{\dag}$ with sensing steps. 
}}
\label{fig:Fig3}
\end{figure}

\newpage{}
\begin{figure}
\includegraphics[width=1.0\textwidth]{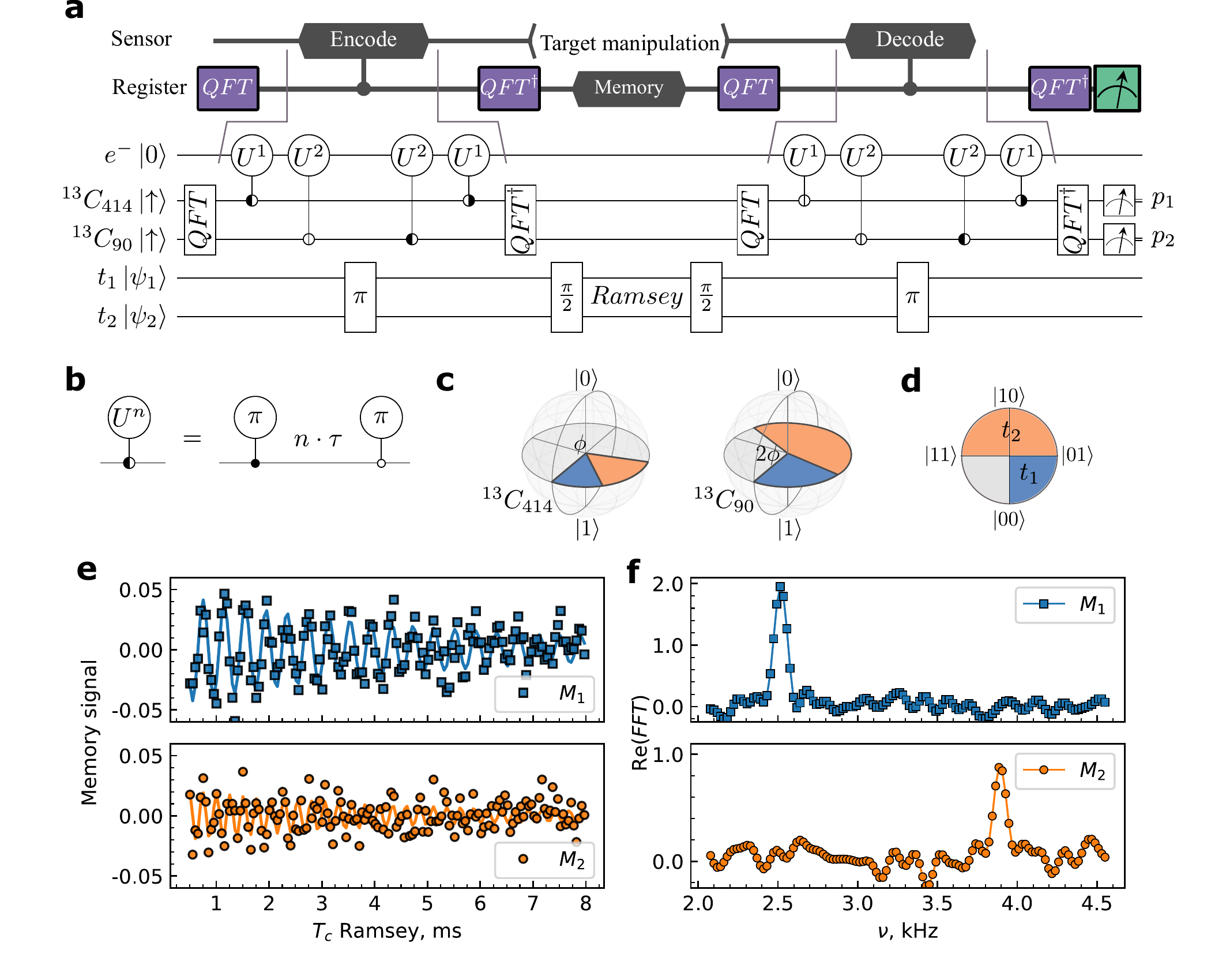} \caption{{
(a) 5-qubit circuit for high resolution spectroscopy. Metrology information is decoded first with a single quantum phase estimation
algorithm step and with a filter to preselect the target spins of interest.
(b) Controlled unitaries $U^n$ consist of two CROT gates with
different states of the control qubit separated by $n\tau$.
(c) Representation of the phase acquired by two register qubits. The phase results from two target spin contributions proportional to their coupling strength and conditional on the target spin state. The net amount of phase on the second qubit is twice larger than on the third in order to match the requirement of the $QFT$ transformation. 
(e) Ramsey measurement on two target spins
as shown in (a). 
After performing an FFT on the data (shown in (f)),
two distinct peaks arise with a linewidth of a 100 Hz allowing to
distinguish between the two target spins.}}
\label{fig:Fig4}
\end{figure}







\end{document}


\preprint{APS/123-QED}

\title{Supplementary information for "QFT for quantum sensing"} 
\author{Vadim Vorobyov$^{1, *,\S}$, Sebastian Zaiser$^{1}$, Nikolas Abt$^{1}$, Jonas Meinel$^{1}$, Durga Dasari$^{1}$, Philipp Neumann$^{1}$ and J$\mathrm{\"o}$rg Wrachtrup$^{1}$}
\affiliation{$^{1}$ 3. Physikalisches Institut, IQST and Centre for Applied Quantum Technologies, and MPI
for Solid State Research, University of Stuttgart, Pfaffenwaldring
57, 70569 Stuttgart, Germany}

\maketitle

\section{Fisher information}
Here we recall and summarize the formalism of Quantum and Classical Fisher information (QFI, CFI)
\subsection{General definitions}
Consider a quantum state $\rho=\rho(a)$ which encodes the unknown parameter $a$.
The logarithmic derivative operator $L_a$ is defined by the equation:
\begin{equation}
    \partial_a \rho = \frac{1}{2}(L_a \rho + \rho L_a)
    \label{L}
\end{equation}
The sum of all derivatives of diagonal elements of density matrix should give zero, because $\mathrm{Tr}(\rho)=1$ for any density matrix, so $\mathrm{Tr}(\partial_a \rho) = 0$, and it follows from equation \ref{L} that:
\begin{equation}
    \mathrm{Tr}(\rho L_a) = 0
    \label{eq:zero}
\end{equation}
The quantum Fisher information could be calculated using the formula:
\begin{equation}
    F_Q := \mathrm{Tr}(\rho L_a^2)
    \label{FQ}
\end{equation}
\subsection{Pure states}
Pure state could be represented as $\rho = \rho^2 = |\psi \rangle \langle \psi |$. By substituting $\rho = \rho ^2$ in the left part of the equation \ref{L} and comparing it to the right part we could obtain the following expression for $L_a$:
\begin{equation}
\begin{split}
     L_a & = 2 \partial_a \rho\\ 
    & = 2 (|\partial_a\psi\rangle \langle \psi| + |\psi \rangle \langle \partial_a\psi |)  
\end{split}
\end{equation}
Substituting $L_a$ to equation \ref{FQ} and placing the expression under $\mathrm{Tr}$ in $\langle \psi | \cdot |\psi \rangle $ we obtain the following calculation for $L^2$ and $QFI$:

\begin{equation}
\begin{split}
    L_a^2 & = 4 \left(|\partial_a\psi\rangle \langle \psi| + |\psi \rangle \langle \partial_a\psi|\right)^2 \\& = \left(|\partial_a\psi\rangle \langle \psi|\right)^2 + |\partial_a\psi\rangle \langle \psi|\psi \rangle \langle \partial_a\psi|\\ & + |\psi \rangle \langle \partial_a\psi|\partial_a\psi\rangle \langle \psi| + \left(|\psi \rangle \langle \partial_a\psi|\right)^2 \\ 
    & = \left(|\partial_a\psi\rangle \langle \psi|\right)^2 + |\partial_a\psi\rangle \langle \partial_a\psi|\\ & + |\psi \rangle \langle \partial_a\psi|\partial_a\psi\rangle \langle \psi| + \left(|\psi \rangle \langle \partial_a\psi|\right)^2
    \end{split}
\end{equation}

\begin{equation}
    \begin{split}
        F_Q & = \langle \psi| \rho L_a^2 |\psi\rangle = \langle \psi | \psi \rangle \langle \psi | L_a^2 |\psi \rangle =  \langle \psi| L_a^2 |\psi\rangle\\
        & = 4(\langle \psi|(|\partial_a \psi \rangle \langle \psi |\partial_a \psi \rangle \langle \psi | + |\psi \rangle \langle \partial_a \psi |\psi \rangle \langle \partial_a \psi | \\
        & +|\partial_a \psi  \rangle \langle \partial_a \psi |+|\psi \rangle \langle\partial_a\psi |\partial_a \psi \rangle \langle \psi |)|\psi \rangle ) \\
        & = 4(\langle \psi |\partial_a \psi \rangle \langle \psi |\partial_a \psi \rangle + \langle \partial_a \psi |\psi \rangle \langle\partial_a\psi |\psi \rangle \\
        & +\langle \psi |\partial_a\psi \rangle \langle \partial_a\psi |\psi \rangle + \langle \partial_a\psi |\partial_a \psi \rangle )\\
        & = 4\left(\langle \partial_a \psi | \partial_a \psi \rangle + (c+c^{\dag})^2 - |c|^2\right)
    \end{split}
\end{equation}
Where $c = \langle \psi |\partial_a \psi \rangle$. By placing $L_a$ into equality \ref{eq:zero} and using $\langle \psi |\cdot |\psi \rangle$ we can show that:
\begin{equation}
\begin{split}
    0 &= \mathrm{Tr}\left( |\psi \rangle \langle \psi |(|\partial_a \psi \rangle \langle \psi | + |\psi \rangle \langle \partial_a \psi |)\right) \\
    & = \langle \psi |\partial_a \psi \rangle + \langle \partial_a \psi | \psi \rangle = c+c^{\dag} = 0.
    \end{split}
\end{equation}
By doing this we obtain the final formula for QFI of a pure state which coincides with the formula 21 from the review by Liu et. al. 2019
\begin{equation}
\begin{split}
        F_Q =& 4(\langle \partial_a \psi | \partial_a \psi \rangle - |\langle \partial_a \psi | \psi \rangle |^2). 
    \end{split}
    \label{eq:fq}
\end{equation}

\subsection{Classical Fisher information}
For results of the measurements we only measured the population of our register states $|b_0,..,b_n\rangle$, hence a classical Fisher information analysis could be made. We used the formula for classical Fisher information: 
\begin{equation}
F_C(\phi)=\sum_{(b_0,...,b_n)}\frac{1}{p(b_0,...,b_n|\phi)}
\Big[\frac{\partial p(b_0,...,b_n|\phi)}{\partial \phi}\Big]^2  
\end{equation}
We have to take a summation over all possible states, which would be $2^n$  combinations in case of $n$ qubits. 
The result of the calculation is a function of phase $\phi$. In the end we estimate the average of $F_C(\phi)$ on a $0 - 2 \pi$ interval.

\subsection{QFI and CFI for QFT, SQL and NOON (Heisenberg)}
To compare the performance of our sensing protocol to other known ones, we calculate the QFI for these three methods for various number $n$ of qubits in use. 
When the qubits are not entangled, an SQL limit is observed, since the system could be decomposed as $|\rho\rangle = \bigotimes_{i=1}^n \rho_i$, so according to the properties of the QFI it is equal to $F_Q = n \cdot F_1$ where $F_1$ is the QFI for single qubit. In that case the final state is $\psi = \otimes_{j=0}^n |0\rangle + e^{i \phi}|1\rangle$, and consequently the $F_q = \sum_{j=0}^{n-1} 1= n$.
If a phase estimation algorithm and $QFT$ are applied to the same initial state with non entangled qubits the phase is written onto the register according to $\phi_i = \phi_0 2^i$, where $\phi_0$ is the phase argument and phase written onto the first qubit. The $QFT^{\dag}$ is applied and the state $|\psi\rangle = QFT^{\dag} |\psi_0\rangle$ is obtained. 
The state after the phase estimation algorithm is equal to $\psi = \otimes_{j=0}^n |0\rangle + e^{i 2^j \phi}|1\rangle$. The estimation of the Fisher information using the equation \ref{eq:fq} applied to that state reads $F_q = \sum_{j=0}^{n-1} 2^{2j}$. Since the $QFT^\dag$ is a unitary transformation which does not depend on $\phi$ it is not affecting $F_q$. 
In the case of the HL or a NOON initial state and phase estimation algorithm the final state is $\psi = \frac{1}{\sqrt{2}}|000..\rangle + e^{i \sum_{j=0}^{n-1} 2^j \phi} |111..\rangle$. The corresponding Fisher information reads as $F_q = (\sum_{j=0}^{n-1} 2^j)^2$.
Utilizing equation \ref{eq:fq} and by calculating $|\partial_{\phi}\psi\rangle$ also numerically using the QuTiP python package \cite{Johansson2013}, we estimate the QFI for $QFT$ at various numbers of qubits $n$ and check that the $QFT$ does not change the value of the Fisher Information. 
\begin{figure}
    \centering
    \includegraphics[width=\columnwidth]{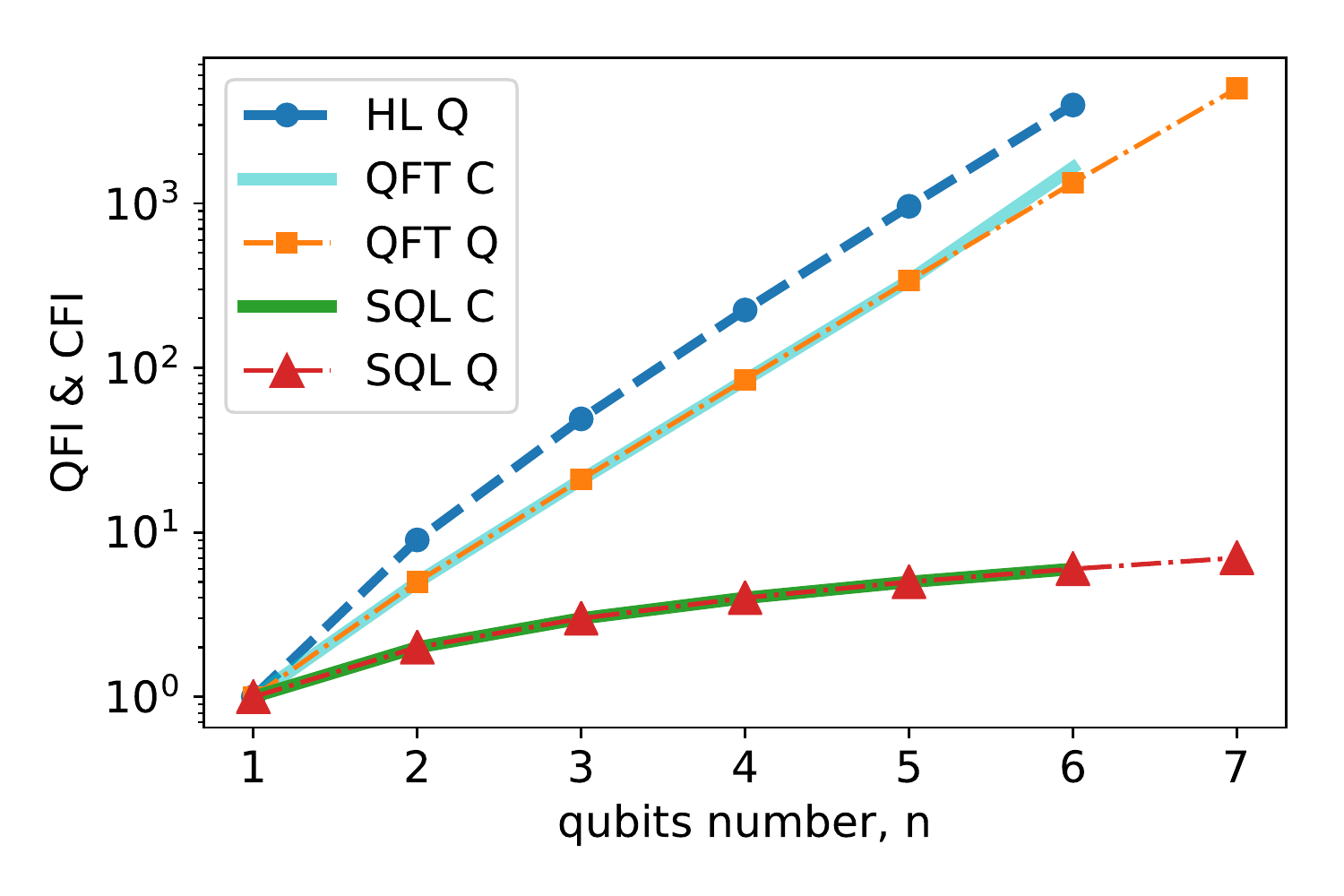}
    \caption{Fisher information estimated for the HL, SQL and QFT protocol. HL, QFT, SQL stands for Heisenberg scaling, phase estimation protocol and conventional scaling respectively. Q and C denotes the quantum and classical Fisher information calculated using the formulas in the text.}
    \label{fig:qf}
\end{figure}
In a real measurement one measures the probability of eigen state of the system, and hence the classical Fisher Information could be used, applied to the classical outputs of the measurements. To check the difference we plot the Quantum and Classical Fisher information of our process (see Fig. \ref{fig:qf}).

\section{QFT circuit for qutrit-qubit system}
\subsection{Conventional QFT}
To design an efficient quantum circuit to perform $QFT$ on our hybrid qutrit-qubit system we generalize the derivation of the $QFT$ circuit made for conventional multi qubit quantum register (see e.g. book by Nielsen, Chuang).
$QFT$ is the quantum analogous of the discrete Fourier transform. It transforms the computational basis (population basis) of the register onto the Fourier basis (phase basis) of the register. In the case of a single qubit a $QFT$ gate is presented by single $H$- Hadamard gate. 
For $n$ qubits the $QFT$ is presented using the following formula:
\begin{equation}
    QFT|j\rangle = \frac{1}{\sqrt{n}} \sum_{k=0}^{n-1} e^{2\pi i j k/n}|k\rangle    \label{eq:conv_qft} 
\end{equation}
\subsection{QFT for general Qudit Register}
By utilizing qubits with more then two states, qudits, ($d>2$) one could gain more computational power while using the same number of physical systems. Thus an extension of the formula \ref{eq:conv_qft} is needed to be used with a general qudit system with dimensions of qudits $d = \{d_1,d_2..d_n\}$ constituting the register. 
Consider a quantum register of $n$ qudits of dimensions $d$ with a total Hilbert space dimension equal to $N = \Pi_{i=1}^{n} d_i$. An integer $k = 0, ... N-1$ could be represented in hybrid base digital representation as: 
\begin{equation}
    k = \{k_1 k_2 .. k_n\} = k_n + \sum_{l=1}^{n-1} k_l \prod_{m=l+1}^n d_m.
    \label{eq:qft_pre1}
\end{equation}
where $k_l = 0,1,...d_l-1$. The base of digital representation is $d$. For example a binary representation is a special case of such a representation. 
By dividing equation \ref{eq:qft_pre1} by the total size of the register $N$ we can obtain the following useful expression:
\begin{equation}
\begin{split}
    \frac{k}{N} &= k_n \prod_{m=1}^n d_m^{-1} + \sum_{l=1}^{n-1}k_l \prod_{m=l+1}^n d_m \prod_{o=1}^n d_o^{-1} \\
    &=\sum_{l=1}^n k_l \prod_{m=1}^l d_l^{-1}.
    \label{eq:qft_pre2}
\end{split}
\end{equation}
The following expression is used to transfer qudit encoding of a state onto a register eigen state state: 
\begin{equation}
j = j_1+ \sum_{l=2}^n j_l \prod_{m=1}^{l-1} d_m  .  
\end{equation}
We note that $j$ is given in a reversed basis, as opposed to $k$. 
In our derivation we will use the following expression, since $e^{2\pi i x}=1, x\in Z$
\begin{equation}
\begin{split}
    &\mathrm{exp}\left(2 \pi i \left[ j \prod_{m=1}^l d_m^{-1}\right]\right) \\
    &=\mathrm{exp}\left( 2\pi i \left[ j_1 \prod_{m=1}^l d_m^{-1} + \sum_{p=2}^n j_p \prod_{o=1}^{p-1}d_o \prod_{m=1}^l d_m^{-1}\right] \right) \\
    & = \mathrm{exp}\left( 2\pi i \left[ j_1 \prod_{m=1}^l d_m^{-1} + \sum_{p=2}^l j_p \prod_{o=1}^{p-1}d_o \prod_{m=1}^l d_m^{-1}\right] \right)\\
    & = \mathrm{exp}\left( 2\pi i \left[ j_1 \prod_{m=1}^l d_m^{-1} + \sum_{p=2}^l j_p \prod_{m=p}^{l} d_m^{-1}\right] \right)\\
    & = \mathrm{exp}\left( 2\pi i \left[ \sum_{p=1}^l j_p \prod_{m=p}^{l} d_m^{-1}\right] \right).\\
\end{split}    
\end{equation}
Now we can rewrite the definition of a $QFT$ for a general multi-qudit register with dimension vector $d$ using  the expression for $k/N$:
\begin{equation}
    \begin{split}
        QFT|j\rangle & = \frac{1}{\sqrt{N}}\sum_{k=0}^{n-1} e^{2\pi i j k /N}|k\rangle \\
        & = \frac{1}{\sqrt{N}} \sum_{k_1=0}^{d_1-1} .. \sum_{k_n}^{d_n-1} e^{2\pi i j \left( \sum_{l=1}^n k_l \prod_{m=1}^{l} d_m^{-1}\right)} |k_1 .. k_n\rangle\\    
        & = \frac{1}{\sqrt{N}} \sum_{k_1=0}^{d_1-1} .. \sum_{k_n}^{d_n-1} \bigotimes_{l=1}^{n} e^{2\pi i j k_l \prod_{m=1}^{l} d_m^{-1}} |k_l\rangle\\
        & = \frac{1}{\sqrt{N}}  \bigotimes_{l=1}^{n} \sum_{k_l=0}^{d_l-1} e^{2\pi i j k_l \prod_{m=1}^{l} d_m^{-1}} |k_l\rangle\\
        & = \frac{1}{\sqrt{N}}  \bigotimes_{l=1}^{n} \sum_{k_l=0}^{d_l-1} e^{2\pi i k_l \sum_{p=1}^l j_p \prod_{m=p}^l d_m^{-1}}  |k_l\rangle\\
        & = \frac{1}{\sqrt{N}}  \bigotimes_{l=1}^{n} \sum_{k_l=0}^{d_l-1} e^{2\pi i k_l 0.j_l..j_2 j_1}  |k_l\rangle .\\
    \end{split}
    \label{eq:general_qft}
\end{equation}
Where in the last line the phase representation is defined with $l = 1,2,...,n$ as :
\begin{equation}
    0.j_l..j_2 j_1 = \frac{j_l}{d_1} + ... + \frac{j_2}{d_1...d_{l-1}} + \frac{j_1}{d_1...d_l}.
\end{equation}

Now a quantum circuit could be derived from that representation for a hybrid qudit based register. Below we provide as an example derivation of a circuit for the single three level system (qutrit) and twelve level system containing of single qutrit and two qubits.
\subsection{QFT for qutrit}
For this case we set in the equation \ref{eq:general_qft} $n=1$ and $d = \{ 3 \}$. 
\begin{equation}
\begin{split}
    QFT_{\{3\}}|j\rangle &= \\
    & = \frac{1}{\sqrt{3}}  \sum_{k_l=0}^{2} e^{2\pi i k_l j_1  /3}  |k\rangle\\
    &=\frac{1}{\sqrt{3}}\left[|0\rangle + e^{\frac{2\pi i j_1}{3}}|1\rangle + e^{\frac{4\pi i j_1}{3}}|2\rangle  \right]\\
    & = C|j\rangle.
    \end{split}
\end{equation}
Where $C$ is the Chrestenson gate \cite{AlRabadi.2002}:
\begin{equation}
  C=\frac{1}{\sqrt{3}}\begin{pmatrix}
    1 & 1 & 1\\
    1 & e^{\frac{2\pi i}{3}} &e^{\frac{4\pi i}{3}}  \\
    1 & e^{\frac{4\pi i}{3}}  & e^{\frac{2\pi i}{3}} .
    \end{pmatrix}   
\end{equation}
It is an analogous to a Hadamard gate to three level system. In our experimental realisation a nitrogen $^{14}N$ represents the three level system. The Chrestenson gate is realized as a two-frequency optimal control pulse. 

\subsection{QFT for 12-level system}
Our 12-level quantum register consists of three nuclear spins ($n=3$) with dimensions $d = \{3,2,2\}$: one qutrit ($d_1=3$) and two qubits ($d_2 = d_3 = 2$). By setting $n=3$ and $d$ into the equation \ref{eq:general_qft} we obtain a $QFT$ expression for that case:
\begin{equation}
    \begin{split}
        QFT_{\{322\}}|j\rangle& = \frac{1}{\sqrt{12}}  \bigotimes_{l=1}^{3} \sum_{k_l=0}^{d_l-1} e^{2\pi i k_l \sum_{p=1}^l j_p \prod_{m=p}^l d_m^{-1}}  |k_l\rangle\\
        & = \frac{1}{\sqrt{3}}\left(|0\rangle + e^{2\pi i\left[ \frac{j_1}{3}\right]}|1\rangle + e^{2\pi i\left[ \frac{2j_1}{3}\right]}|2\rangle  \right)\\
        &\otimes \frac{1}{\sqrt{2}}\left(|0\rangle + e^{2\pi i \left[ \frac{j_1}{6}+\frac{j_2}{2}\right]}|1\rangle \right)\\
        &\otimes \frac{1}{\sqrt{2}}\left(|0\rangle + e^{2\pi i \left[\frac{j_1}{12}+\frac{j_2}{4}+\frac{j_3}{2}\right]}|1\rangle \right).\\
    \end{split}
    \label{eq:qft_12Levels}
\end{equation}
From the expression \ref{eq:qft_12Levels} we can derive a gate sequence using the following considerations. First the state $j_3$ is processed. It has to acquire eventually the phase $2\pi i \left[\frac{j_1}{12}+\frac{j_2}{4}+\frac{j_3}{2}\right]$. By first applying a Hadamard gate $H$, we create a state \begin{equation}
    \psi = H|j_3\rangle = \frac{1}{\sqrt{2}}\left(|0\rangle + e^{2\pi i \left[ \frac{j_3}{2}\right]}|1\rangle\right).
\end{equation}
By applying a controlled phase rotation $Z_{\pi/2}$ after that conditional on $j_2$ (second qubit) we add a phase $i \pi/2 j_2  =2 \pi  i j_2 /4$. In the same manner phase rotation $Z_{\pi/6}$ conditional on qutrit state $j_1$ adds residual required phase to the state. For the second qubit the procedure is analogous, the Hadamard gate and now the phase rotations $Z_{\pi/3}$ conditioned on qutrit state are applied. The operation on the qutrit doesn't depend on other spins and hence analogously to a single qutrit a Chrestenson gate $C$ is performed. 
\begin{figure}[h]
    \centering
    \includegraphics{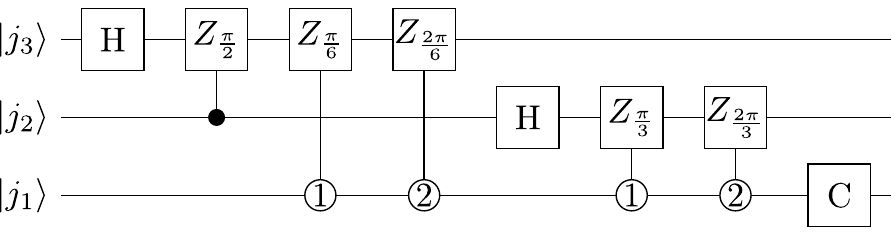}
    \caption{QFT circuit for 12 level register. The state of the register is $|j_1 j_2 j_3\rangle$, with dimensions of the qudits $d=\{3,2,2\}$}
    \label{fig:qft12}
\end{figure}

\section{Experimental methods}
\subsection{Setup}
The experimental setup consists of a homebuilt confocal microscope, a permanent magnet on a motorized stage for the creation of the external magnetic field $B_0$ aligned with the NV quantization axis and equipment for electron and nuclear spin manipulation. The setup operates at ambient conditions, i.e. room temperature and atmospheric pressure and is used exclusively to work with single NV centers and described in details elsewhere \cite{Waldherr2014}. 

\subsection{Sample}

The diamond sample is embedded into a sapphire wafer of 2 mm thickness and a diameter of 50 mm. The sapphire wafer is mounted on a 3-axis piezoelectric scanner with a travel range of $100 \, \mu m x 100 \, \mu m x 25 \, \mu m$ and nanometer resolution. The NV center and its surrounding nuclear spins which are used in current work as a quantum register are described in previous work \cite{Waldherr2014}.
In our experiments we use a single electron spin as a sensor, which is associated with the nitrogen-vacancy center (NV) in diamond. 
The electron spin has a long coherence time at ambient conditions despite its large coupling to external fields and consequently a high intrinsic sensitivity. 
The register is formed by three nuclear spins, one $^{14}N$nuclear spin which has spin $I=1$ (qutrit) and two $^{13}C$
which have spin $I=1/2$ (qubits) forming an effective 12-level system and allows to run a general set of quantum operations including the QFT algorithm. 
Additionally, our register allows for projective readout of nuclear spins and thus efficient extraction of sensing information \cite{Neumann.2010b, Waldherr2014}. Together electron spin sensor and nuclear spin register form a quantum enhanced sensor. 
The NV center has several hyperfine coupled nuclear spins in the environment having much longer coherence times as compared to the electron spin but having a low susceptibility to the environment. 
The co-processor is formed using the strongly coupled ($A_{||}\approx 2.16$   $\mathrm{MHz}$) $^{14}N$ nuclear spin of the NV center, and two $^{13}C$ nuclear spins (labelled as $^{13}C_{1}$ and $^{13}C_{2}$) with hyperfine coupling along the quantization axis of the NV center $A_{zz}\approx414$ $\mathrm{kHz}$ and $A_{zz}\approx 89$  $\mathrm{kHz}$ respectively.
Due to these large splitting's we can resolve all the twelve lines in the ODMR spectrum of the NV center \cite{Waldherr2014}.
This marks the key step for our protocol, as the digitization is encoded as populations among the 12-levels, and the phase digitized with a resolution of $\phi/12$. The individual resolution of these levels set the limit on our precision. 
Using strongly coupled nuclear spins, as memory qubits allows them to be individually addressed and read out. 

\subsection{Nuclear register initialization}


An important step is to initialize the memory qubits.
This is done by exploiting deterministic initialization, using a SWAP operations with the polarized electron spin of NV \cite{Waldherr2014}.
The initialization fidelity in such case determined by the fidelity of electron spin optical initialization and a SWAP gate. To further increase the fidelity of initialization we perform projective measurement and post-selection of the results based on initialization projective measurement. The nuclear spins could be readout with single shot technique with fidelities $F_r$ of 95.8\% ($^{14}N$), 96.9\% ($^{13}C_414$) and 99.6\% ($^{13}C_90$). The initialization fidelities larger 99\% could be achieved \cite{Waldherr2014}.


\subsection{Direct quantum memory phase writing}
For writing-in and storing the sensed phase information in the memory register, we use the direct quantum memory access (DQMA) \cite{Pfender2017}, where the sensor gets entangled with the memory qubit, and
a projection of the sensor spin in a chosen basis initializes the register spins in targeted states. 
This method though not general enough when compared to the SWAP operation, makes a natural choice
while implementing Quantum Phase Estimation with $iQFT$ (\textit{QPE}) as it reduces the computational cost of the overall implementation. In the DQMA method, the circuit also starts with initialized qubits $|\psi\rangle = |0\rangle_e\bigotimes_{i=1}^{n} |0\rangle_{n_i}$, and the memory qubit is mapped onto the equatorial plane of the Bloch sphere using a Hadamard gate $H$ creating a $|psi\rangle = |0\rangle_e \bigotimes_{i=1}^n \frac{1}{\sqrt{2}}\left(|0\rangle + |1\rangle\right)_{n_i} $, where $n$ is number of qubits, $e$ and $n_i$ stands for electron and nuclear spin indices respectively.
Subsequently, a CNOT gate implemented as a selective microwave pulse entangles sensor and a memory $|\psi\rangle= \frac{1}{\sqrt{2}}\left(|0_e0_t\rangle + |1_e1_t\rangle\right)\bigotimes_{i=1, i\ne t}^n \frac{1}{\sqrt{2}}\left(|0\rangle + |1\rangle\right)_{n_i} $ 
This configuration exhibits the susceptibility to the environment of the electron spin and imprints the phase acquired by the electron spin onto the target nuclear memory. 
After disentangling with a second selective $\pi$ pulse, the electron is in the eigenstate and the quantum information is stored on the memory qubit \cite{Pfender2017}.

\subsection{Quantum register control gates}
Besides reading and writing to the memory, information processing among the co-processor units is necessary. With rf and microwave manipulation NOT gates on the qubits, CNOT gates between sensor and memory qubits
(which is important for writing) as well as non-local among the memory qubits, which are crucial to run a quantum co-processor with any algorithm, can be implemented. We thus have a complete set of operations to perform
any operation between the sensor and the co-processor. 
For the $QFT$, the relevant gates are Hadamard gates and controlled phase gates. 
The Hadamard gate is experimentally implemented using $\pi_y/2$ pulses. 
Though it does not have the conventional form of the Hadamard matrix, it is similar up to a sign change of two matrix elements and preserves the mapping from population to phase. 
The key element, the controlled-phase gate is implemented using the hyperfine interaction
between the nuclear spins and the electron spin, and the gate is performed conditioned on the electron spin being in the $m_{s}=|1\rangle$. 
As we use two $^{13}C$ spins in our register, the method of working in $m_{s}=0$ and using the hyperfine interaction will not work out anymore since the carbon nuclear spins are not distinguishable in $m_{s}=0$, so we have to switch to $m_{s}=-1$. To realize a controlled phase gate, the optimal control method is applied. 
The optimal control way of implementing the CPHASE gates is very convenient since we only have to know the free evolution Hamiltonian and specify the task we want to optimize. 
The two and three qubit $iQFT$ implemented using optimal control pulses show good experiment with theory using ideal pulses (see Fig.2 in the main text)

\subsection{Sensing of external AC field}
\begin{figure}
    \centering
    \includegraphics[width=\columnwidth]{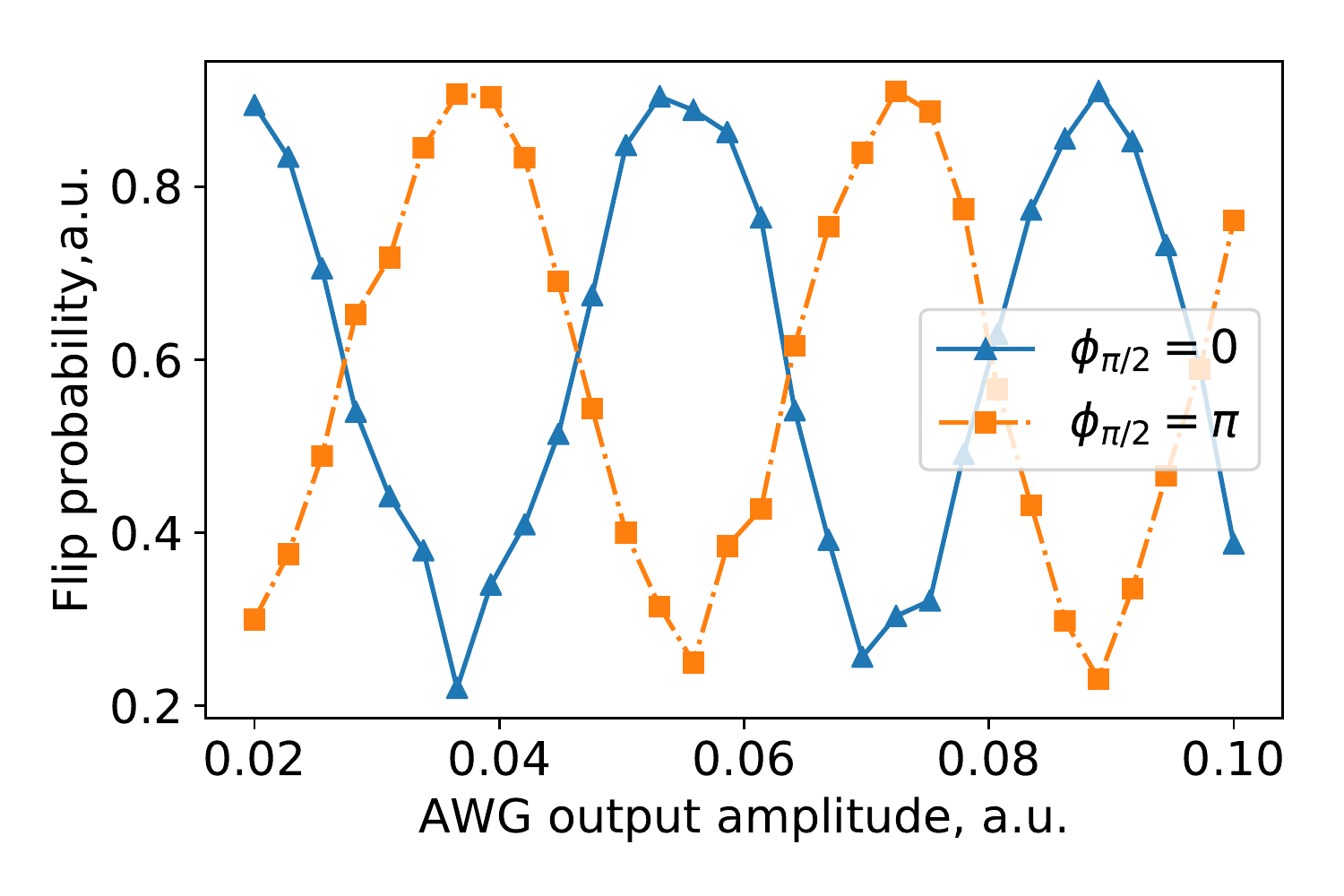}
    \caption{Sensing of external field by sensor and one memory}
    \label{fig:ac field}
\end{figure}

To perform measurement of an oscillating magnetic field we utilize a conventional dynamical decoupling sequence $XY-N$ (DD) combined with direct memory access \cite{Pfender2017}. During application of DD we apply externally an oscillating field at the frequency given by the filter function generated with the DD. To regulate the acquired phase during the sequence due to the artificial field we vary the number of pulses of the DD and hence total interrogation according to the phase estimation algorithm and Fig.3b in the main text. In the DD the $\pi$ pulses have length $100$ $\mathrm{ns}$ and are not selective to particular nuclear spin subspace. The robustness of the sequence is accomplished by using pulse sequence of $XY$ type. (See Fig. \ref{fig:ac field}).

\subsection{Correlation spectroscopy using QFT}

In accordance with the Fig. 4a of the main text in the first sensing step, depending on the initial state of the target spins, the register acquires a phase as follows $\phi_{LSQ}=2\cdot 2\tau (A_1 I_1+ A_2 I_2)$, and $\phi_{MSQ}=2\cdot \tau (A_1 I_1+ A_2 I_2)$. 
and possible outputs of the memory after the first sensing steps are listed in table \ref{tab:phases}.  
The second (\textit{correlating}) sensing step acquires a phase with a negative sign, such that no change in the signal during the correlation time leads to a zero net phase. Consequently, the flip of $t_2$ spin during the correlation time give respectively $2\pi$ on LSB and $\pi$ on MSB. We note that it does not affect the LSB but fully flips the MSB. The flip of the 6 kHz target spin results in a $\pi$ phase shift on the LSB and $\pm\pi/2$ on the MSB. Here, there is a full flip of the LSB but also an additional phase on the MSB. This additional phase is corrected by the final $QFT^{\dag}$. Firstly, the $QFT^{\dag}$ transfers LSB onto the population state, and crucially, depending on it's state, performs the controlled phase gate, which corrects this $\pm \pi/2$ phase on MSB if the LSB acquired $\pi$ flip.

\begin{table}[h]
    \centering
    \begin{tabular}{|c|c|c|}
         \hline
         $\ket{12 kHz} \otimes \ket{6 kHz}$ & LSB & MSB \\
         \hline
         \hline
         $\ket{\uparrow\uparrow}$& $3\pi/2$ &$3\pi/4$\\
                  \hline
         $\ket{\uparrow\downarrow}$& $\pi/2$ &$\pi/4$\\
                  \hline
         $\ket{\downarrow\uparrow}$& $-\pi/2$ &$-\pi/4$\\
                  \hline
         $\ket{\downarrow\downarrow}$& $-3\pi/2$ &$-3\pi/4$\\
                  \hline
    \end{tabular}
    \caption{The phase acquired on the register after single sensing step}
    \label{tab:phases}
\end{table}

\subsection{Single Memory example}
In a simple case where a single memory qubit is used, we digitize the phase into single bits. 
We start with both the sensor spin and the memory spin initialized.
The first two gates that are applied, a nuclear $\pi/2$(which is a QFT gate for 1 qubit register) and a CROT gate, entangle memory and sensor spin. 
This leads to the entanglement of the memory and sensing qubit, thus allowing the total state to acquire phase. 
During the sensing time $\tau$ the sensor-memory state, $\ket{11}$, is acquiring a phase due to external magnetic fields and after a second CROT gate the state is disentangled and the phase information is imprinted on the memory spin state,
mimicking the action of a phase gate $U$.
Higher powers $U^{n}$ can then be realized by accumulating phase over longer time scales $n\tau$. 
The phase retains on the memory spin for a correlation time $T$, before the sensor spin and memory spin get entangled and disentangled again similar to the initial step for a second write process. 
The second write process leads to another phase accumulation on the state $\ket{10}$ and the phase difference $\delta\phi$ of these two accumulated phases on the subspace is then measured by mapping it into population. 
The phase which is accumulated during any sensing time $\tau$ is given by $\phi=\gamma_{e}B_{z}\tau$, where $\gamma_{e}$ is the gyromagnetic ratio of the electron spin sensor and $B_{z}$ the magnetic field being the source of phase accumulation. 
If the environment does not change during $T$, the correlation of phases $\phi_{1}$ and $\phi_{2}$ will yield no signal ($\delta\phi=0$), since they are both equal.
However, if the nuclear target spins in the environment are flipped on purpose by an rf $\pi$-pulse, the phase on the memory spin will then be given by $\delta\phi=\mp2\pi A_{zz}\tau$, where $A_{zz}$ is the hyperfine interaction strength of the target nuclear spin with the sensor spin. 
To achieve the highest susceptibility to a spin which is coupled with $A_{zz}$, the sensing time must be matched to achieve $\delta\phi=\pi$, i.e. $\tau=1/2A_{zz}$.
This is a flip of the memory which is the maximum change possible and hence the maximum signal.

\subsection{Quantum Phase estimation 2 Memory example}
For a quick overview on the working principle of QFT based phase estimation that we employ in our experiment, let us consider a simple example with two qubits as the memory register and one qubit that acts as
a sensor. The qubit which is collecting the most phase will be called
the fast oscillating or the least significant bit (LSB). The other
qubit with the lowest amount of phase will be called the slow oscillating
or most significant bit (MSB). Depending on the state of the sensor
qubit the memory spins acquire a phase of $\phi$ (LSB) and $2\phi$
(MSB), respectively. Consequently, the phase of the LSB is transformed
into population via another Hadamard gate and depending on this outcome,
a phase of $\pi/2$ is subtracted from the phase $\phi$ of the MSB
and this shifted phase is then transformed into population. The circuit
implementation of this is performed with the memory qubits initialized
in $\ket{00}$, followed by Hadamard gates that put them into a superposition
state $[(\ket{0}+\ket{1})/\sqrt{2}]$. The memory qubits
perform controlled phase gates on the sensor qubit, thereby registering
the phase acquired by it through its interaction with the target qubit.
Following this we perform inverse QFT i.e., a Hadamard gate on the
LSB, and a C-ROT gate on the MSB followed by a Hadamard gate. The
C-ROT gate subtracts a phase of $\pi/2$ on MSB conditioned on the
state of LSB. Finally, we measure all the four states of the memory
register. The measurement results would be $[00,01,10,11]$, when
the phase $\phi$ takes values {[}0, $\pi/2$, $3\pi/2$, $2\pi${]}.
We would like to note that in the case where no iQFT is performed
the memory register could only distinguish two phases. Hence, using
two memory qubits, we are sensitive to phase changes in units of $\pi/2$,
i.e., we could subdivide the total phase $\pi$ into four equal parts
that can be clearly distinguished. Hence to increase the phase sensitivity all we need is to increase the memory register such that we are become sensitive to phase changes in units of $\pi/2^{(N-1)}$. 

\subsection{Phase initialization of the quantum register}
To mimick the acquisition of the phase $\phi$ into the register we phase shift the $\pi/2$ pulses and $C$ gates by a phase $\phi$.

\subsection{Purity conservation}
\begin{figure}
    \centering
    \includegraphics[width=\columnwidth]{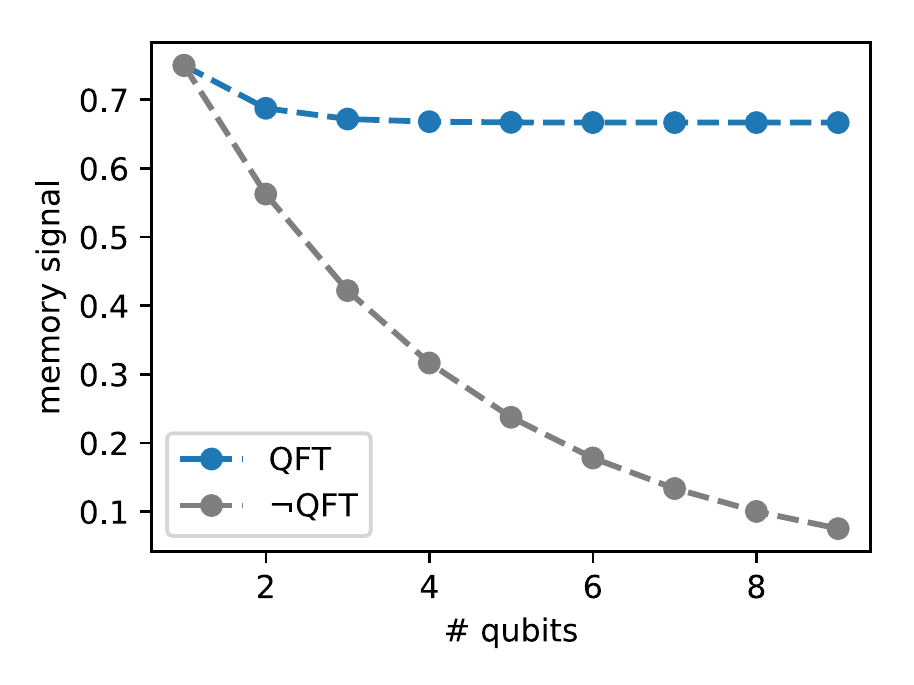}
    \caption{The simulated average of the purity left in the quantum register after the $T_2$ decay. The QFT denotes the application of the $QFT^\dag$ algorithm to the register, and $-QFT$ denotes the application of conventional local $H$ gates.}
    \label{fig:purity}
\end{figure}

An important benefit which $QFT$ brings to the methods of quantum correlation sensing is preservation of multi-qubit quantum register purity during the correlation time. 
Imagine we have an in-situ correlation sensing protocol \cite{Laraoui.2013, Zaiser.2016}.
After the first sensing state, we transfer the acquired phase to the population basis of the multi-qubit register and store it for $T_1$ time, exceeding the $T_2$ time of the nuclear memory.
In that case the coherent components of the register qubits states will decay, due to the decoherence. 
As seen from Fig. 2 in the main text, in case of the most straightforward conversion of the phase to the population basis via application of local Hadamard gates (e.g. using $\pi/2$ pulses), the spread of the population among many register states leads to the loss of the purity, since at once there are many individual qubits which are not in the $I_z$ eigen state. 
On the other hand, when a $QFT$ is applied, the population is at most spread among the neighbouring register states or even concentrated inside single eigen state of the register. On average it means that most of the qubits stay at the $I_z$ eigen states and hence do not loose their purity, except 1 or 2 qubits, whose state is not an eigen state. As it is seen in Fig. \ref{fig:purity} the purity is reducing with number of used qubits dramatically without usage of the $QFT$ algorithm. That emphasizes the additional importance of using QFT in extracting the quantum information and storing it for long time for correlation.
\bibliography{references} 
\bibliographystyle{ieeetr}
\clearpage